\documentclass[aps,pra,groupedaddress,floats,reprint]{revtex4-2}

\usepackage[english] {babel}
\usepackage{amsfonts,graphicx,soul,natbib,mathrsfs,dsfont}

\usepackage[colorlinks=true]{hyperref}

\usepackage{epstopdf}
\usepackage{xcolor}
\usepackage{soul}
\usepackage{physics}
\usepackage{graphics}
\usepackage{comment}

\renewcommand{\vec}[1]{\textnormal{\boldmath$#1$}}

\usepackage{tikz}
\usepackage{amscd}
\usepackage[inline]{enumitem}
\usepackage[
    left=1.9cm,
    right=1.9cm,
    top=1.5cm,
    bottom=2cm,
    bindingoffset=0cm,
]{geometry}
\usepackage{amsmath, nicefrac}
\newcommand{\sign}{\text{sign}}
\usepackage{cancel}
\usepackage{physics}

\usepackage[normalem]{ulem}

\usepackage{dcolumn}
\usepackage{bm}
\begin{document}

\title{Twisted charged particles in the uniform magnetic field with broken symmetry}

\author{N. V. Filina}%
\affiliation{School of Physics and Engineering,
ITMO University, St. Petersburg, Russia 197101}%

\author{S. S. Baturin}%
\email{s.s.baturin@gmail.com}%
\affiliation{School of Physics and Engineering,
ITMO University, St. Petersburg, Russia 197101}%

\date{\today}
\begin{abstract}
We present a theoretical description of
charged particles with nonzero projection of the orbital angular momentum (OAM) in
a uniform magnetic field with broken axial symmetry.  
The wave functions we find naturally account for the asymmetry of the magnetic field at the entrance of the solenoid through the continuous parameter and are a generalization of the Laguerre-Gauss states commonly used to describe twisted charged particles. We analyze the asymmetric Hamiltonian from an algebraic point of view and show how the OAM projection of the twisted state is modified by symmetry breaking. We provide analytical frameworks for properties of the asymmetric states, such as energy, RMS size, and Cazimir invariant, and discuss advantages of the proposed description.                
            
\end{abstract}

\maketitle

The symmetry group of a system is the fundamental element that completely defines the properties of the problem under consideration. 
Since Noether's discovery \cite{Noether1918}, symmetry arguments have been widely used in theoretical analysis
of physical problems. The main idea is the generators of the system's symmetry group that are closely related to the conserved properties of this
system. For example, the projection of the orbital angular momentum (OAM) along the $z$-axis remains constant in systems with cylindrical symmetry. Recent theoretical \cite{Bliokh2007,Bliokh2012,Barnett2017,ct4,Armen} and experimental studies \cite{Exp0,Exp01,Exp02,Exp1,Exp2} reveal that electrons in free space or in a constant magnetic field can
carry a non-zero OAM projection. Such quantum states are called \textit{twisted electrons}. The OAM projection is conserved only and only if the system is axially symmetric and, according to the
Noether's theorem, may vanish
once this symmetry is broken. In free space, the symmetry can be guaranteed with a high degree of accuracy, but in a solenoid with a constant magnetic field the situation is different.

The first solution to this
problem was derived by Landau \cite{Landau:1930aa,Landau} in a fully asymmetric (Landau) gauge, 
widely known as Landau states. At first sight, Landau states have nothing to do with twisted electrons. Recently, however, it has been shown
that in the case of the axily symmetric system \cite{Bliokh2012}, Landau states can have a singularity in the phase of their
wave function, which implies a
twisted particle \cite{Allen1999,Bliokh2012}. It is well known that all observables must be gauge independent, so to detect experimentally the difference between these two solutions the physical symmetry of the problem must manifest itself. These symmetry is defined by the distribution of currents in space and consequently by the structure of the magnetic field that is well defined and unique in contrast to the vector potential. In the case of the magnetic solenoid, the structure of the transverse components of the magnetic field at the entrance defines this symmetry. 


Indeed, lets consider the most simple case when magnetic field appear abruptly at some point in space $B_z=B_0\theta(z)$, where $\theta(z)$ is the Heaviside step-function. The magnetic field must be divergence-free and thus we have
\begin{align}
\vec{\nabla}\cdot\vec{B}=\partial_x B_x+\partial_y B_y+B_0 \delta(z)=0.
\end{align}  
We immediately observe that along with the step in the longitudinal magnetic field there must be a delta discontinuity in the transverse magnetic field. Assuming the linearity of both $B_x$ and $B_y$ one may chose the magnetic field in the form 
\begin{align}
B_x=-(1-\beta) x B_0 \delta(z),~~B_y=-\beta y B_0 \delta(z)
\end{align}
that accounts for the asymmetry of the transverse magnetic field.
Above $\beta$ - is the symmetry parameter. Thus, the vector of the magnetic field near the $z$ axis is
\begin{align}
\label{eq:magF}
\vec{B}^{T}=B_0 \left\{-(1-\beta) x \delta(z);-\beta y \delta(z); 
 \theta(z) \right\}.
\end{align}
The field above corresponds to a simplified model of a magnetic field at the entrance of the elliptical solenoid with $x$ and $y$ directed along the semi-major and semi-minor axes. Since we are interested in the field near the axis, we will decompose the vector potential up to the second order into spatial coordinates near the origin.
The most general form of the vector potential, which is quadratic in spatial coordinates and corresponds to the magnetic field Eq.\eqref{eq:magF}, has the form
\begin{align}
\label{eq:vpgen}
    \vec{A}=B_0\left(\begin{matrix} \theta(z)\left[a x + b y \right] \\ \theta(z)\left[(1+b) x + c y \right] \\ \delta(z)\left[\frac{a}{2} x^2 + d x y +\frac{c}{2} y^2\right] \end{matrix}\right).
\end{align}
%
with the additional condition
\begin{align}
d-b=\beta
\end{align}
that follows from the connection $\vec{B}=\vec{\nabla}\times\vec{A}$. Note that the freedom of choice in the arbitrary constants $a,b,c,d$ reflects the gauge freedom in the definition of the vector potential.

We use relativistic units $\hbar = c = 1, e < 0$ in our further analysis.

Consider an interaction of the electron with the magnetic field in minimal coupling. The electron's kinetic momentum is
\begin{align}
\hat{\vec{p}}_{\mathrm{kin}}=\hat{\vec{p}}-e\vec{A}.
\end{align} 
We note that in general all three components of the kinetic momentum depend on the transverse coordinates as follows from Eq.\eqref{eq:vpgen} and the problem is three dimensional. 

It is convenient to choose the gauge potential $\chi$ such that only two of the three components of the kinetic momentum depend on the transverse coordinates
\begin{align}
\label{eq:GP}
\chi=-B_0\theta(z)\left(\frac{a}{2} x^2 + d x y +\frac{c}{2} y^2 \right).
\end{align}
Performing gauge transformation 
\begin{align}
\vec{A}'=\vec{A}+\nabla\cdot \chi
\end{align}
we have
\begin{align}
\label{eq:vpot}
(\vec{A}')^T=B_0\left\{-\beta \theta(z)y; (1-\beta) \theta(z) x, 0\right\}.
\end{align}
%
The form of the vector potential given above now explicitly takes into account the real asymmetry of the magnetic field at the boundary between the free space and the magnetic field regions, and allows one to separate the variables and to analyze the transverse eigenstates of the electron for the case $z>0$ on the basis of the simple two-dimensional model. We note that the potential given by Eq.\eqref{eq:vpot} satisfies the Coulomb gauge condition $\vec{\nabla}\cdot \vec{A}'=0$ and under the restrictions of the model, the gauge potential Eq.\eqref{eq:GP} is the only way to satisfy the Coulomb gauge condition. Thus, we simply use the Coulomb gauge, which completely fixes the vector potential in the considered approach to the solution. Note that a different gauge will result in a different setup, but will leave the results of the analysis intact.

Next, we focus on the transverse part of the nonrelativistic Schr\"{o}dinger equation for a massive charged particle in a magnetic field, which 
has the form 
\begin{align}
\label{eq:mainSch}
    i\partial_t \psi=\hat H_\perp \psi,~~~\hat{H}_\perp=\frac{\left[\hat{\vec{p}}_\perp-e \vec{A}'_\perp \right]^2}{2 m}. 
\end{align}
It can be shown that the evolution of the transverse part of the wave function of a
non-relativistic point particle for $z>0$ is reduced to Eq.\eqref{eq:mainSch} \cite{Baturin}. Moreover, the same type of equation will appear in the Foldy-Wouthuysen 
representation under the paraxial approximation after a proper substitution of $t \to z$ and dropping the spin \cite{SilenkoFW}. We adhere to the formulation of the problem
given in Eq.\eqref{eq:mainSch}, but note that the analysis and results apply directly to the relativistic case with minor modifications.

Inserting the vector potential given by Eq.\eqref{eq:vpot} into 
Eq.\eqref{eq:mainSch}, we get for $z>0$
\begin{align}
    \hat{H}_\perp = \; &\frac{\vec{\hat{p}}^2_\perp}{2 m_e}- \sign(e) 2 \omega \left[- \beta \hat{p}_x \hat{y} + (1 - \beta) \hat{p}_y \hat{x}\right]\nonumber \\
    &+ 2 m_e \omega^2 \left[\beta^2 \hat{y}^2 + (1 - \beta)^2 \hat{x}^2\right],
\end{align}
where we have introduced the Larmor frequency:
\begin{align}
    \omega = \frac{|e| B}{2 m_e}.
\end{align}
To separate the symmetric and antisymmetric parts of 
$\hat{H}_\perp$, we make a canonical transformation:
\begin{align}
    \hat{\widetilde{x}} &= \sqrt{2 (1 - \beta)} \hat{x} \;\;\;\;\;\;\;\;\;\; \hat{\widetilde{y}} = \sqrt{2 \beta} \hat{y} \nonumber\\
    \hat{\widetilde{p}}_x &= \frac{\hat{p}_x}{\sqrt{2 (1 - \beta)}} \;\;\;\;\;\;\;\;\;\;\; \hat{\widetilde{p}}_y = \frac{\hat{p}_y}{\sqrt{2 \beta}}
\end{align}
Then, we rearrange the terms, and the Hamiltonian takes the form (here, we omit the waves for brevity of notation):
\begin{align}
\label{eq:ham}
    \hat{H}_\perp = 
    &\left[ \left( \frac{\hat{p}^2_x}{2 m_e} + \frac{\hat{p}^2_y}{2 m_e}\right) + \frac{m_e \omega^2 (\hat{x}^2 + \hat{y}^2)}{2} \right] \nonumber \\
    &+ (1 - 2\beta) \left[ \left( \frac{\hat{p}^2_x}{2 m_e} - \frac{\hat{p}^2_y}{2 m_e}\right) + \frac{m_e \omega^2 (\hat{x}^2 - \hat{y}^2)}{2}\right] \nonumber\\
    &- \sign(e) 2 \omega \sqrt{\beta (1 - \beta)} \hat{L}_z
\end{align}
For 
convenience, we introduce the following operators 
\begin{align}
\label{eq:operdef}
    \hat{H}_s &= - \frac{1}{2 m_e} \left(\frac{\partial^2}{\partial x^2} + \frac{\partial^2}{\partial y^2} \right) + \frac{m_e \omega^2 (x^2 + y^2)}{2}, \nonumber\\
    \hat{H}_1 &= - \frac{1}{2 m_e} \left(\frac{\partial^2}{\partial x^2} - \frac{\partial^2}{\partial y^2} \right) + \frac{m_e \omega^2 (x^2 - y^2)}{2}, \nonumber \\
    \hat{H}_2 &= - \frac{1}{m_e} \frac{\partial^2}{\partial x \partial y} + m_e \omega^2 x y, \\\nonumber
    \hat{H}_3 &= \omega \hat{L}_z = - i \omega \left(x \frac{\partial}{\partial y} - y \frac{\partial}{\partial x} \right). 
\end{align}

In the new notations, the Hamiltonian 
\eqref{eq:ham} can be expressed as 
\begin{align}
\label{eq:hamsimple}
    \hat{H}_\perp &= \hat{H}_s + \hat{H}_{as}, \\ \nonumber
    \hat{H}_{as}(\alpha) &= - \sign(e) \left[ \cos(2\alpha) \hat{H}_1 + \sin(2 \alpha) \hat{H}_3 \right],
\end{align}
where we have introduced a new symmetry parameter $\beta = \sin^2 \widetilde{\alpha}$ and $\widetilde{\alpha} = \frac{\pi}{4} + \sign(e) \left( \frac{\pi}{4} - \alpha \right)$ and $\hat{H}_s$ is defined in Eq.\eqref{eq:operdef}.  

For 
the stationary problem, when the Schr\"{o}dinger equation Eq.\eqref{eq:mainSch} is reduced to
\begin{align}
\label{eq:stSch}
    \left[\hat{H}_s+\hat{H}_{as}(\alpha)\right]\psi =\epsilon \psi,
\end{align}
Eq.\eqref{eq:stSch} can be solved \textit{exactly}, and the corresponding solutions generalize the known Hermite-Gauss (HG) and Laguerre-Gauss (LG) states to the \textit{asymmetric Landau states} (ALS), i.e. 
states corresponding to an intermediate symmetry defined by the parameter $\alpha$ in the Hamiltonian 
\eqref{eq:hamsimple}. 

The complete set of orthogonal eigenfunctions of $\hat{H}_\perp$ is given by Hermite-Laguerre-Gauss (HLG) 
functions \cite{EGAbramochkin_2004} that are a special case of Ince-Gaussian mode \cite{Bandres:04} and reads (see the Supplementary Information)
\begin{align}
    &\psi_{n,m}(x,y,\alpha)=\mathcal{G}^{N}_{n,m}\left(\tilde x,\tilde y| \alpha \right), \\
    &\tilde x= \frac{x}{\rho_H}, ~~~ \tilde{y}=\frac{y}{\rho_H}. \nonumber
\end{align}
with the transverse energy of a 
state given by
\begin{align}
\label{eq:enrg}
    &\epsilon^-=2\omega\left(n+\frac{1}{2}\right), ~~ \sign(e)<0, \nonumber\\
    &\epsilon^+=2\omega\left(m+\frac{1}{2}\right), ~~ \sign(e)>0.
\end{align}
Above, we have introduced the Landau radius as
\begin{align}
    \rho_H=\sqrt{\frac{2}{m_e \omega}}.
\end{align}
HLG functions were first discovered in quantum optics as a special class of solutions to the paraxial wave equation that are invariant under astigmatic influence. It is not surprising, though, that the same functions naturally appear in the Landau problem, as the Schr\"{o}dinger equation is very similar to the paraxial wave equation \cite{SilenkoG,SilenkoFW}. For 
charged particles, however, 
the astigmatism stems 
from the asymmetry of the transverse magnetic field at the boundary.

An 
HLG function 
has two natural limiting cases
$\alpha=0$ and $\alpha=\pi/4$. The former corresponds to the completely asymmetric HG eigenstates given by Hermite polynomials with zero projection of the OAM on the $z$ axis. 
\begin{align}
    \psi_{n,m}(x,y,0)&=\\ \nonumber &\frac{(-i)^me^{-\frac{x^2+y^2}{\rho_H^2}}\mathcal{H}_{n}\left(\sqrt{2}\frac{x}{\rho_H}\right)\mathcal{H}_{m}\left(\sqrt{2}\frac{y}{\rho_H}\right)}{\rho_H\sqrt{\pi 2^{n+m-1}n!m!}}.
\end{align}
Here, $\mathcal{H}_{n}(x)$ 
is a 
Hermite polynomial of order $n$.

\begin{figure}[t]
    \centering
     \includegraphics[width=0.5\textwidth]{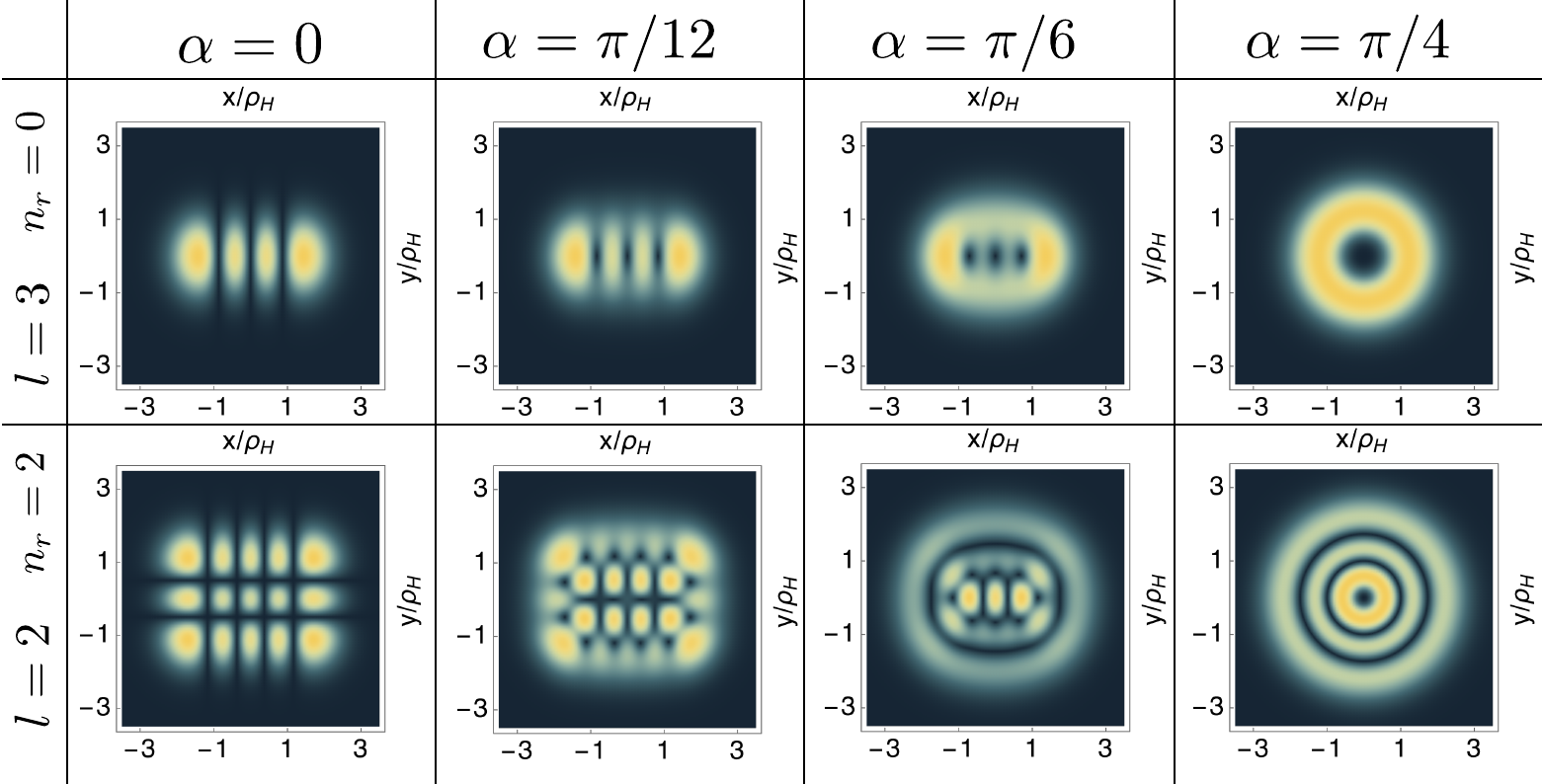}
    \caption{Probability density for 
    asymmetric Landau states for different values of the symmetry parameter $\alpha$. Top 
    row corresponds to 
    $n_r=0$ and $l=3$, bottom row to 
    $n_r=2$ and $l=2$.}
    \label{Fig:1}
\end{figure}

The opposite case of $\alpha=\pi/4$ corresponds to the symmetric eigenfunctions commonly known as \textit{twisted states}, or the LG states with a 
defined projection of the OAM. For $n\leq m$, we have (the case of $n\geq m$ looks similar except for the reversed sign of the OAM)
\begin{align}
\label{eq:LG}
    &\psi_{n,m}\left(x,y,\pi/4 \right)=\frac{(-1)^n2^m n!}{\rho_H\sqrt{\pi 2^{n+m-1}n!m!}}\left(\frac{\sqrt{x^2+y^2}}{\rho_H}\right)^{|m-n|} \nonumber\\  &\mathcal{L}_{n}^{|m-n|}\left(2\frac{x^2+y^2}{\rho_H^2}\right)e^{-\frac{x^2+y^2}{\rho_H^2}-i(m-n)\arctan(y/x)},
\end{align}
where $\mathcal{L}_{n}^{|m-n|}(x)$ is a 
generalized Laguerre polynomial.
A detailed description of HLG functions and their properties can be found in numerous publications by Abramochkin and coauthors \cite{EGAbramochkin_2004,AbramBook,Abram_last}.  
It is 
convenient to introduce another set of 
quantum numbers: 
the radial quantum number $n_r$ and 
the eigenvalue of the $z$-projection of the OAM $l$, which usually characterizes the twisted state:  
\begin{align}
 \label{eq:qnTW}   
    l &= n - m, \nonumber\\
    n_r &=
        \frac{n+m-|l|}{2}. 
\end{align}
With Eq.\eqref{eq:qnTW} and Eq.\eqref{eq:LG}, we obtain the 
familiar form of an 
LG state in 
cylindrical coordinates
\begin{align}
    &\psi_{n_r,l}\left(r,\phi,\pi/4 \right)\propto \left(\frac{r}{\rho_H}\right)^{|l|}\mathcal{L}_{n_r}^{|l|}\left(\frac{2r^2}{\rho_H^2}\right)e^{-\frac{r^2}{\rho_H^2}+il\phi}.
\end{align}
To illustrate the dependence of 
HLG modes (or ALS) on the normalized transverse coordinates, we plot their probability density distributions for different values of $\alpha$ and fixed values of quantum numbers $n_r$ and $l$ in Fig.\ref{Fig:1}; there, we can see that the ALS are highly sensitive 
to the symmetry. 
From Fig.\ref{Fig:1}, we can also conclude 
that the asymmetry visually reveals the value of the OAM projection.

To proceed, we compare various parameters of an 
ALS with the corresponding values of a 
common twisted LG state. 
The energy of an 
ALS is obtained from Eq.\eqref{eq:enrg} with Eq.\eqref{eq:qnTW} as follows
\begin{align}
\label{eq:ALSenrg}
    \epsilon=\omega\left[2n_r+|l|- \sign(e)l+1\right]
\end{align}
which is exactly the same as for an 
LG state. 

We can evaluate different mean values using 
the recurrence 
relation for the functions $\mathcal{G}^{N}_{n,m}(x/\rho_H,y/\rho_H|\alpha)$ and their orthogonality property. The mean square radius of an 
ALS in terms of quantum numbers 
\eqref{eq:qnTW} is
\begin{align}
    \langle r^2\rangle &= \iint \limits_{\mathds{R}^2} (x^2 + y^2) |\mathcal{G}^N_{n,m}(x/\rho_H,y/\rho_H | \alpha)|^2 dx dy, \nonumber\\\langle r^2\rangle&=\frac{\rho_H^2}{2} (2n_r + |l| + 1),
\end{align}
which, once 
again, coincides with the same average calculated for 
the LG states. 

Obviously, 
the $\hat{L}_z$ operator does not commute $[\hat{L}_z,\hat{H}_\perp]\neq0$ with the Hamiltonian 
\eqref{eq:hamsimple}; consequently, ALS are not 
eigenfunctions of $\hat{L}_z$, except for a fully symmetric case of $\alpha=\pi/4$. However, the mean value of $\hat{L}_z$ for 
the ALS states has a simple form
\begin{align}
\label{eq:Lzm}
    \langle \hat{L}_z\rangle &= -i\iint \limits_{\mathds{R}^2} \overline{\mathcal{G}^N_{n,m}(x,y | \alpha)} x \partial_y\mathcal{G}^N_{n,m}(x,y | \alpha) dx dy+  \nonumber \\&i \iint \limits_{\mathds{R}^2} \overline{\mathcal{G}^N_{n,m}(x,y | \alpha)} y \partial_x \mathcal{G}^N_{n,m}(x,y | \alpha) dx dy, \nonumber \\
    \langle \hat{L}_z\rangle &=l \sin{2\alpha}.
\end{align}
As expected, the OAM projection on the $z$ axis vanishes in the fully asymmetric case of $\alpha=0$. In the limiting case, $\hat{H}_{as}(\pi/4)=-\sign(e)\omega\hat{L}_z$ and $[\hat{H}_\perp,\hat{L}_z ]=0$. Thus, 
it seems reasonable 
to assume that 
when $\alpha\neq \pi/4$, the second integral of motion is equal to just 
$\hat{H}_{as}(\alpha)$. It is easy to check (see the Supplementary Information
) that, indeed, 
\begin{align}
    \forall \alpha \;\; [\hat{H}_\perp,\hat{H}_{as}(\alpha)]=0.
\end{align}
Furthermore, 
using the properties of the $\mathcal{G}^{N}_{n,m}(x/\rho_H,y/\rho_H|\alpha)$ function once again, we get
\begin{align}
\label{eq:aseigen}
    \hat{H}_{as}(\alpha)\psi_{n,m}(x,y,\alpha)=-\sign(e)\omega l \psi_{n,m}(x,y,\alpha).
\end{align}
Strikingly, we see that the OAM projection of 
an LG state in the asymmetric case is actually an eigenvalue of a more complex operator, which is conserved under symmetry breaking. 

To gain further insight, we first note the following equivalence
\begin{align}
    \hat{H}_2\psi(x,y,\alpha)=-i\omega \partial_\alpha \psi(x,y,\alpha).
\end{align}
and recall that $\hat{H}_3\propto -i\partial_\varphi$. Note that both operators are generators of rotations, since $\varphi$ and $\alpha$ are periodic. 
Direct evaluation of the commutators 
\begin{align}
\label{eq:algb}
    \left[ \hat{H}_i , \hat{H}_j \right] &= 2 i \omega \varepsilon_{i j k} \hat{H}_k,~~i,j,k\in\{1,2,3\},
\end{align}
shows the exact equivalence of the operator algebra $\hat{H}_i$ with the SO(3) algebra of pseudo angular momentum operators $\hat{\mathfrak{L}}_i=\hat{H}_i/2\omega$. Here, $\varepsilon_{i j k}$ is a 
totally antisymmetric Levi-Civita tensor. 

The symmetric part of the Hamiltonian commutes with all three pseudo angular momentum operators
\begin{align}
\label{eq:shwH}
    \left[ \hat{H}_s , \hat{H}_i \right] = 0,~~i\in \{1,2,3\}.
\end{align}
and reminds of 
a general Schwinger model \cite{Schwinger} of the two-dimensional harmonic oscillator, where the full Hamiltonian consists of the isotropic part (
in the present case, $\hat{H}_s$) and the 
sum of the three coupling pseudo angular momentum operators $\propto\hat{H}_{1,2,3}$. 

\begin{figure}[t]
    \centering
     \includegraphics[width=0.5\textwidth]{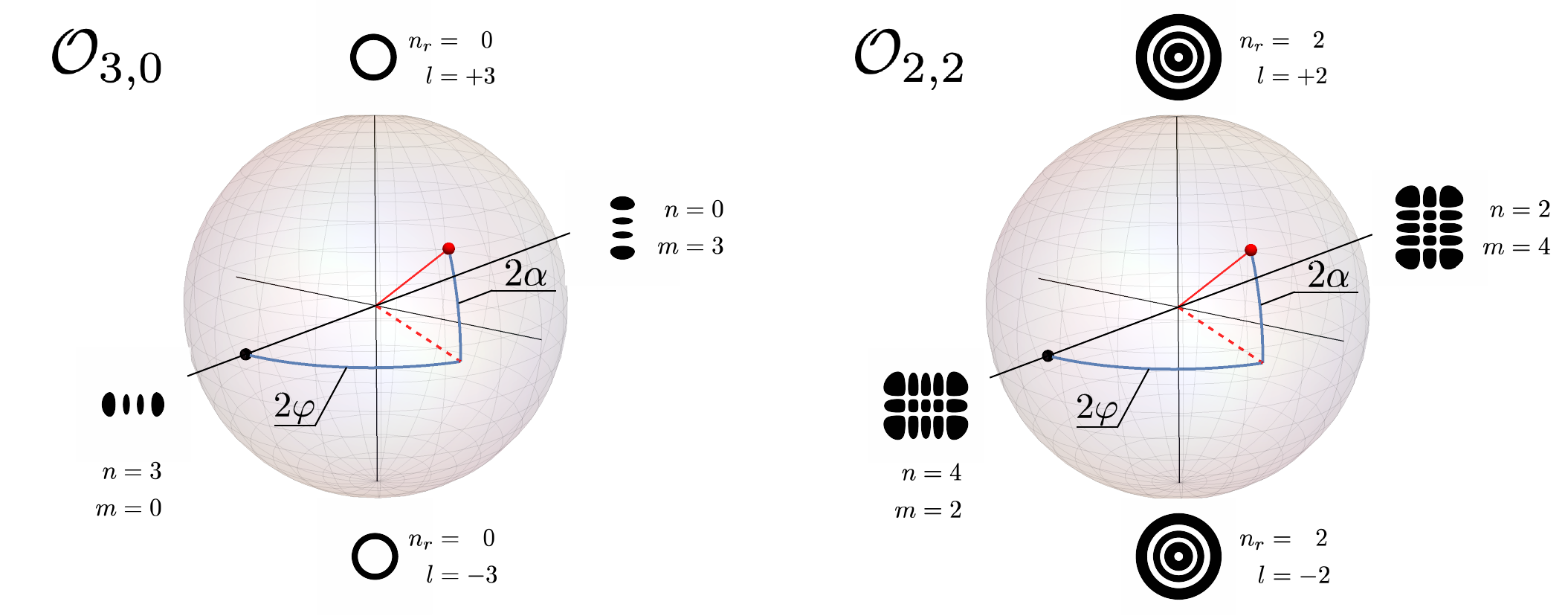}
    \caption{Orbital Poincaré spheres of 
    asymmetric Landau states.}
    \label{Fig:3}
\end{figure}

Indeed, under a clockwise rotation $\hat{R}(-\varphi)$ of the $XY$-plane Hamiltonian, Eq.\eqref{eq:hamsimple} is transformed as follows
\begin{align}
    &\hat{R}\hat{H}_s\hat{R}^{-1}=\hat{H}_s, \\
    &\hat{R}\hat{H}_{as}\hat{R}^{-1}=- \sign(e)2\omega \vec{n} \hat{\vec{\mathfrak{L}}},
\end{align}
where
\begin{align}
\vec{n}^{T}=\left(\cos 2\varphi \cos 2\alpha,\sin 2\varphi \cos 2\alpha,\sin 2\alpha \right)    
\end{align}
is a unit vector in the space of three orthogonal axis that correspond to the operators $\hat{\mathfrak{L}}_{1,2,3}$; $(\hat{\vec{\mathfrak{L}}})^{T}\equiv\frac{1}{2\omega}\left(\hat{H}_1,\hat{H}_2,\hat{H}_3 \right)$ 
is the vector of pseudo angular momentum operator (see Fig.\ref{Fig:3}).

Thus, the most general Hamiltonian for a twisted asymmetric state has the form
\begin{align}
\label{eq:Sch}
    \hat{H}=\hat{H}_s- \sign(e)2\omega \vec{n} \hat{\vec{\mathfrak{L}}},
\end{align}
and the \textit{generalized} ALS $\psi^{sch}$, which 
is an eigenstate of the Schwinger Hamiltonian, Eq.\eqref{eq:Sch}, can be expressed as a simple rotation of the ALS $\psi$ and has the following form:

\begin{align}
\label{eq:genALS}
    &\psi^{sch}_{n,m}(x,y,\alpha)= \hat{R}\psi_{n,m}(x,y,\alpha)= \\ \nonumber
    &\psi_{n,m}(x\cos\varphi+y\sin\varphi,-x \sin\varphi+y\cos\varphi,\alpha). 
\end{align}

We recognize that the unit vector $\vec{n}$ has a meaning of the \textit{spin 
axis} for the ALS on the orbital Poincaré sphere \cite{Calvo:05,Habraken:10,RevModPhys2022}. 
The \textit{spin axis} can either be directly observed in the case of LG states as the $z$-projection of the OAM, or completely hidden from observation, as in the case of HG states. The latter can 
also be seen from the mean value of $\hat{L}_z$ given by Eq.\eqref{eq:Lzm}. 

To proceed further, we note that under a similarity transformation $\hat{R}(-\varphi)$, Eq.\eqref{eq:aseigen} is transformed to
\begin{align}
    \vec{n} \hat{\vec{\mathfrak{L}}}\psi^{sch}_{n,m}(x,y,\alpha)=m_l\psi^{sch}_{n,m}(x,y,\alpha),
\end{align}
and we can see 
that $\psi^{sch}_{n,m}(x,y,\alpha)$ is an eigenfunction of the operator of the projection of the pseudo angular momentum onto the \textit{spin 
axis} with an eigenvalue equal to half the OAM values:
\begin{align}
\label{eq:mlnum}
    m_l=\frac{l}{2}.
\end{align}

We note that the coefficient 1/2 comes from the charge of the algebra given by Eq.\eqref{eq:algb} and indicates 
that the pseudo angular momentum operators $\hat{\mathfrak{L}}_i$ are operators of the three orthogonal projections of the pseudo spin.

Next, we recall that the SO(3) algebra has a Casimir invariant, which can be calculated 
as the square of the modulus of the pseudo angular momentum vector 
\begin{align}
    \hat{\mathfrak{K}}=|\hat{\vec{\mathfrak{L}}}|^2=\frac{1}{4\omega^2}\sum\limits_{i}\hat{H}_i^2.
\end{align}
The Casimir operator commutes with the Hamiltonian Eq.\eqref{eq:Sch}, so the functions $\psi^{sch}_{n,m}(x,y, \alpha)$ are eigenfunctions of the operator $\hat{\mathfrak{K}}$. Using the definitions of Eq.\eqref{eq:operdef}, we derive 
the explicit form of the Casimir operator and get (see the Supplementary Information)
\begin{align}
    \hat{\mathfrak{K}}=\frac{\hat{H}_s^2}{4\omega^2}-\frac{1}{4}.
\end{align}

Note that eigenvalues of $\hat{\mathfrak{K}}$ can be found by applying this operator to 
any equivalent set of functions that are eigenvectors of $\hat{H}_s$. The simplest choice is the set of HG states corresponding to a set of $\psi_{n,m}(x,y,0)$. Evaluating the action of $\hat{\mathfrak{K}}$ on $\psi_{n,m}(x,y,0)$, we have
\begin{align}
\label{eq:Cas}
    &\hat{\mathfrak{K}}\psi_{n,m}(x,y,0)=\\ \nonumber &\frac{1}{4}\left[(n+m+1)^2-1\right]\psi_{n,m}(x,y,0).
\end{align}
On the other hand, the eigenvalue of the Casimir operator can 
be expressed through 
the pseudo total angular momentum quantum number $j$ and is equal to $j(j+1)$. With Eq
.\eqref{eq:Cas}, we obtain
\begin{align}
    j=\frac{n+m}{2},
\end{align}
or, if expressed through the quantum numbers of the twisted state introduced in 
Eq.\eqref{eq:qnTW},
\begin{align}
\label{eq:jnum}
    j=n_r+\frac{|l|}{2}.
\end{align}
Consequently, the eigenvalue of the Casimir operator can be expressed as 
\begin{align}
    j(j+1)=\left(n_r+\frac{|l|}{2}\right)\left(n_r+\frac{|l|}{2}+1\right).
\end{align}

As we have 
$j$ given by Eq.\eqref{eq:jnum} and $m_l$ given by Eq.\eqref{eq:mlnum}, we can express generalized ALS through 
HG states with the help of Wigner functions $D_{m_l',m_l}^{j}$ 
\cite{Varshalovich} as follows
\begin{align}
\label{eq:genALS2}
    &\psi^{sch}_{j+m_l,j-m_l}(x,y,\alpha)= \\ \nonumber
    &\sum\limits_{m_l'=-j}^{j}D_{m_l',m_l}^{j}(A,B,C)\psi_{j+m'_l,j-m'_l}(x,y,0). 
\end{align}
Here, $A,B,C$ 
are the Euler angles 
defined by the Hamiltonian through 
the following relations \cite{AbramBook}:

\begin{align}
   &e^ {\left(-i\frac{A+C}{2}\right)} \cos \frac{B}{2}=-\sin \varphi \cos \alpha +i \cos \varphi \sin \alpha, \\ \nonumber
   &e^{\left(i\frac{A-C}{2}\right)} \sin \frac{B}{2}=\sin \varphi \sin \alpha+i \cos \varphi \cos \alpha.
\end{align}  

For example, in the case of an HLG state, $\varphi=0$, $A=C=0$, $B=\pi/2-2\alpha$, 
and vector $\vec{n}^{T}=(\cos 2\alpha,0,\sin 2\alpha)$. Substituting 
these expressions into Eq.\eqref{eq:genALS} and using the general Hamiltonian Eq.\eqref{eq:Sch}, we  
recover the definition of the function $\mathcal{G}_{n,m}(x,y|\alpha)$ and the initial Hamiltonian Eq.\eqref{eq:hamsimple}.

We note that the motion of an 
ALS state along the orbital Poincare sphere is related not only to the change in $\alpha$ and $\varphi$, but also to the phase. 

According to the results of \cite{Calvo:05} for a 
similar optical problem, once set to motion ALS states should aquire a non-trivial Berry phase \cite{Berry1}, which is proportional to the OAM projection of the ALS. 
\begin{align}
    \Phi_B=i\oint \langle \psi^{sch}_{n,m}| \vec{\nabla}_{\phi,\theta} | \psi^{sch}_{n,m} \rangle d\vec \Gamma = -\frac{l}{2} {\Omega}.
\end{align}
Above, $\Omega$ is the solid angle enclosed by the path on the orbital Poincaré sphere, $\phi=2\varphi$, and $\theta=\pi/2-2\alpha$. This fact reveals the topological nature of the OAM \cite{Habraken:10,GeomPhase2010}, 
universal for both electrons and photons. For the case of the electrons the motion of the ALS is triggered if the symmetry of the initial twisted electron state in a free space at the point $z=0$ differs from the symmetry of the transverse magnetic field at the entrance. The motion has an oscillatory nature and is essentially a rotation of the incoming twisted state about the \textit{spin axis} of the Hamiltonian.

In conclusion, we should recall 
that the Hamiltonian given by Eq.\eqref{eq:ham} is written in normalized coordinates. Consequently, to return 
to the real coordinates, the 
inverse canonical transformation is required. 
This results in a stretch of the corresponding probability density, but preserves the topology and structure of the ALS.

The results reported in the present paper are directly related to the transformation of modes in quantum optics with mode converters \cite{Conv1993,Allen1}, but here, the 
underlying physics is different. In optics, the asymmetry comes from the astigmatism of the optical focusing channel and is related to the symmetry of the lenses, whereas in the Landau problem considered here, the symmetry is defined by the symmetry of the magnetic field at the entrance of the solenoid. Consequently, it is the symmetry of the magnetic field at the boundary that reveals itself in the visual pattern of the electron probability density. Moreover, if this symmetry is completely broken, then the OAM projection becomes zero, 
but it can always be recovered once the symmetry is restored.

The generalized ALS given by Eq.\eqref{eq:genALS} and Eq.\eqref{eq:genALS2} is the most general stationary solution that explicitly accounts for the symmetry of the considered Landau problem and continuously bridges two extreme cases of the HG and LG states. In full analogy to the common Landau states the axisymmetric case, which give rise to the class of nonstationary Landau states \cite{Silenko2021}, generalized ALS can be extended to a class of non-stationary solutions once combined with the Ermakov mapping \cite{Filina2023} and ideas of the quantum Arnold transformation \cite{QAT1}.

\begin{acknowledgments}
The work is funded by the Russian Science Foundation and the St.\ Petersburg Science Foundation, project № 22-22-20062, https://rscf.ru/project/22-22-20062/. 
The authors thank Igor Chestnov and Ivan Terekhov for useful discussions and suggestions and Lidiya Pogorelskaya for careful reading of the manuscript.
\end{acknowledgments}

\begin{widetext}
\begin{center}
\textbf{{Supplementary materials for the \\ "Twisted charged particles in the uniform magnetic field with broken symmetry".}}
\end{center}

\section{Definition of the HLG functions}
By the definition \cite{EGAbramochkin_2004,AbramBook} HLG function $\mathcal{G}_{n,m}(x,y|\alpha)$ reads
\begin{align}
    \mathcal{G}_{n,m}(x,y|\alpha)=e^{-x^2-y^2}\sum\limits_{k=0}^{n+m}i^k \cos^{n-k}(\alpha) \sin^{m-k}(\alpha) P_{k}^{(n-k,m-k)}(-\cos 2\alpha)H_{n+m-k}(\sqrt{2}x)H_{k}(\sqrt{2}y).
\end{align}
Here $H$ - is the Hermite polynomial and $P$ is the Jacobi polynomial. Normalization factor (square of the $L^2$ norm) for the $\mathcal{G}_{n,m}(x,y|\alpha)$ reads
\begin{align}
    ||\mathcal{G}_{n,m}(x,y|\alpha)||^2=\pi 2^{n+m-1}n!m!.
\end{align}
Consequently, normalized HLG functions $\mathcal{G}^N_{n,m}(x,y|\alpha)$ are introduced as
\begin{align}
    \mathcal{G}^N_{n,m}(x,y|\alpha)=\frac{\mathcal{G}_{n,m}(x,y|\alpha)}{\sqrt{\pi 2^{n+m-1}n!m!}}.
\end{align}
Normalised HLG functions can be defined through the Wigner $D_{m,m'}^{j}$-functions \cite{Varshalovich}
\begin{align}
    \mathcal{G}^N_{j+m,j-m}(x,y|\alpha)=\sum\limits_{m'=-j}^j D_{m',m}^j(0,\frac{\pi}{2}-2\alpha,0) \frac{(-i)^me^{-\frac{x^2+y^2}{\rho_H^2}}\mathcal{H}_{j+m'}\left(\sqrt{2}\frac{x}{\rho_H}\right)\mathcal{H}_{j-m'}\left(\sqrt{2}\frac{y}{\rho_H}\right)}{\rho_H\sqrt{\pi 2^{2j-1}(j+m')!(j-m')!}}.
\end{align}
\section{Evaluation of the commutators}
In the main text the following operators were introduced
\begin{align}
\label{eq:operdef}
    \hat{H}_s &= - \frac{1}{2 m_e} \left(\frac{\partial^2}{\partial x^2} + \frac{\partial^2}{\partial y^2} \right) + \frac{m_e \omega^2 (x^2 + y^2)}{2}, \nonumber\\
    \hat{H}_1 &= - \frac{1}{2 m_e} \left(\frac{\partial^2}{\partial x^2} - \frac{\partial^2}{\partial y^2} \right) + \frac{m_e \omega^2 (x^2 - y^2)}{2}, \nonumber \\
    \hat{H}_2 &= - \frac{1}{m_e} \frac{\partial^2}{\partial x \partial y} + m_e \omega^2 x y, \\\nonumber
    \hat{H}_3 &= \omega \hat{L}_z = - i \omega \left(x \frac{\partial}{\partial y} - y \frac{\partial}{\partial x} \right). 
\end{align}
First, we evaluate commutators of the type $\left[\hat{H}_s,\hat{H}_i\right]$ , with $i\in \{1,2,3\}$:
\begin{align}
    \left[\hat{H}_s,\hat{H}_1 \right]
    = - \frac{\omega^2}{4} \left( 4 x \frac{\partial}{\partial x} + 2 - 4 y \frac{\partial}{\partial y} - 2 \right) + \frac{\omega^2}{4} \left( 4 x \frac{\partial}{\partial x} + 2 - 4 y \frac{\partial}{\partial y} - 2 \right) &= 0,
\end{align}

\begin{align}
    \left[\hat{H}_s, \hat{H}_2 \right]
    = - \frac{\omega^2}{2} \left( 2 y \frac{\partial}{\partial x} + 2 x \frac{\partial}{\partial y} \right) + \frac{\omega^2}{2} \left( 2 x \frac{\partial}{\partial y} + 2 y \frac{\partial}{\partial x} \right) &= 0
\end{align}

\begin{align}
    \left[\hat{H}_s, \hat{H}_3 \right]
    = \frac{i \omega}{2 m_e} \left( 2 \frac{\partial^2}{\partial x \partial y} - 2 \frac{\partial^2}{\partial x \partial y} \right) - i \omega \frac{m_e \omega^2}{2} \left( - 2 x y + 2 x y \right) &= 0.
\end{align}
Next, we evaluate commutators of the type $\left[\hat{H}_i,\hat{H}_j\right]$ , with $i,j\in \{1,2,3\}$
\begin{align}
\label{eq:a1}
    \left[\hat{H}_1, \hat{H}_2 \right]  
    =  \frac{i \omega}{2 m_e} \left( 2 \frac{\partial^2}{\partial x \partial y} + 2 \frac{\partial^2}{\partial x \partial y} \right) - i \omega \frac{m_e \omega^2}{2} \left( 2 x y + 2 x y \right) 
    = - 2 i \omega \hat{H}_2,
\end{align}
\begin{align}
\label{eq:a2}
    \left[\hat{H}_3, \hat{H}_2 \right] 
    = \frac{i \omega}{m_e} \left( \frac{\partial^2}{\partial x^2} - \frac{\partial^2}{\partial y^2} \right) - i \omega m_e \omega^2 \left( x^2 - y^2 \right) 
    = - 2 i \omega \hat{H}_1,
\end{align}
\begin{align}
\label{eq:a3}
    \left[\hat{H}_2, \hat{H}_1 \right] 
    = - \frac{\omega^2}{2} \left( 2 x \frac{\partial}{\partial y} - 2 y \frac{\partial}{\partial x} \right) + \frac{\omega^2}{2} \left( 2 y \frac{\partial}{\partial x} - 2 x \frac{\partial}{\partial y} \right) 
    = - 2 i \omega \hat{H}_3.
\end{align}
Combining Eq.\eqref{eq:a1}, Eq.\eqref{eq:a2} and Eq.\eqref{eq:a3} we get
\begin{align}
\label{eq:algb}
    \left[ \hat{H}_i , \hat{H}_j \right] &= 2 i \omega \varepsilon_{i j k} \hat{H}_k,~~i,j,k\in\{1,2,3\}.
\end{align}
Where $\varepsilon_{i j k}$ is the totally antisymmetric Levi-Civita tensor. 

\section{Casimir invariant}

By the definition the Casimir invariant can be calculated as

\begin{align}
    \hat{\mathfrak{K}} = |\hat{\vec{\mathfrak{L}}}|^2 = \frac{1}{4\omega^2}\sum\limits_{i}\hat{H}_i^2.
\end{align}
Below we evaluate explicitly $\hat{H}^2_i$:
\begin{align}
    \hat{H}^2_1 = \; &\frac{1}{4 m^2_e} \left( \frac{\partial^4}{\partial x^4} \; \uuline{- \; 2 \frac{\partial^4}{\partial x^2 \partial y^2}} + \frac{\partial^4}{\partial y^4}\right) + \frac{m^2_e \omega^4}{4} \left( x^4 \; \uuline{- \; 2 x^2 y^2} + y^4 \right) - \nonumber \\
    &- \frac{\omega^2}{4} \left( 2 x^2 \frac{\partial^2}{\partial x^2} + 4 x \frac{\partial}{\partial x} + 2 + 2 y^2 \frac{\partial^2}{\partial y^2} + 4 y \frac{\partial}{\partial y} + 2 \; \uline{- \; 2 x^2 \frac{\partial^2}{\partial y^2} - 2 y^2 \frac{\partial^2}{\partial x^2}} \right), \\
    \hat{H}^2_2 = \; &\uuline{\frac{1}{m^2_e} \frac{\partial^4}{\partial x^2 \partial y^2}} \; \uuline{+ \; m^2_e \omega^4 x^2 y^2} - \omega^2 \left( \uwave{2 x y \frac{\partial^2}{\partial x \partial y} + y \frac{\partial}{\partial y} + x \frac{\partial}{\partial x}} + 1 \right), \\
    \hat{H}^2_3 = \; &- \omega^2 \left( \uline{x^2 \frac{\partial^2}{\partial y^2} + y^2 \frac{\partial^2}{\partial x^2}} \; \uwave{- \; x \frac{\partial}{\partial x} - y \frac{\partial}{\partial y} - 2 x y \frac{\partial^2}{\partial x \partial y}} \right).
\end{align}
Finally, we add everything up and arrive at the final expression for the Casimir invariant
\begin{align}
    \hat{\mathfrak{K}} = \;&\frac{1}{4 \omega^2} \Bigg[ \frac{1}{4 m^2_e} \left( \frac{\partial^4}{\partial x^4} + 2 \frac{\partial^4}{\partial x^2 \partial y^2} + \frac{\partial^4}{\partial y^4}\right) + \frac{m^2_e \omega^4}{4} \left( x^4 + 2 x^2 y^2 + y^4 \right) - \nonumber \\
    &- \frac{\omega^2}{4} \left( 2 x^2 \frac{\partial^2}{\partial x^2} + 4 x \frac{\partial}{\partial x} + 2 + 2 y^2 \frac{\partial^2}{\partial y^2} + 4 y \frac{\partial}{\partial y} + 2 + 2 x^2 \frac{\partial^2}{\partial y^2} + 2 y^2 \frac{\partial^2}{\partial x^2} \right) - \omega^2 \Bigg] = \nonumber \\
    &= \frac{1}{4 \omega^2} \left[ - \frac{1}{2 m_e} \left(\frac{\partial^2}{\partial x^2} + \frac{\partial^2}{\partial y^2} \right) + \frac{m_e \omega^2 (x^2 + y^2)}{2} \right]^2 - \frac{1}{4} = \frac{\hat{H}_s^2}{4\omega^2}-\frac{1}{4}.
\end{align}

\end{widetext}

\bibliography{references}

\end{document}